\documentclass[twocolumn,showpacs,preprintnumbers,amsmath,amssymb]{revtex4}

\usepackage{graphicx}
\usepackage{dcolumn}
\usepackage{bm}

\begin{document}

\preprint{PRESAT-7802}

\title{First-principles study on field evaporation for silicon atom on Si(001) surface}

\author{Tomoya Ono}
\affiliation{Research Center for Ultra-Precision Science and Technology, Osaka University, Suita, Osaka 565-0871, Japan}

\author{Kikuji Hirose}
\affiliation{
Department of Precision Science and Technology, Osaka University, Suita, Osaka 565-0871, Japan}

\date{\today}

\begin{abstract}
The simulations of field-evaporation processes for silicon atoms on various Si(001) surfaces are implemented using the first-principles calculations based on the real-space finite-difference method. We find that the atoms which locate on atomically flat Si(001) surfaces and at step edges are easily removed by applying external electric field, and the threshold value of the external electric field for evaporation of atoms on atomically flat Si(001) surfaces, which is predicted between 3.0 and 3.5 V/\AA, is in agreement with the experimental data of 3.8 V/\AA. In this situation, the local field around an evaporating atom does not play a crucial role. This result is instead interpreted in terms of the bond strength between an evaporating atom and surface.
\end{abstract}

\maketitle
\section{Introduction}
In recent years, scanning tunneling microscopy experiments in the atomic-scale manipulation of material surfaces have attracted great attention due to its potential to create artificial surface nanostructures. It is now possible to remove adsorbed atoms and molecules from surfaces or to deposit them on surfaces by applying voltage pulses \cite{a}. Field evaporation \cite{tsong} and deposition are thermally activated processes in which rate constants can be parameterized according to the Arrhenius formula \cite{Arrhenius}. In addition, the electric field between probing tip and surface plays a crucial role during these processes, since the activation energy for field evaporation is considered to depend on this electric field. In the past, phenomenological models have been used to evaluate the activation energy for field evaporation of surface atoms \cite{tsong}. However, in order to understand this phenomenon properly from a microscopic point of view, it is mandatory to calculate self-consistently both the electronic charge distribution and the resulting electrostatic field.

In this paper, we carry out first-principles molecular-dynamics simulations to explore the activation energies and threshold values of external electric fields for the evaporation of surface atoms from Si(001) surfaces. Five years ago, Kawai {\it et al.} \cite{kawai} determined the adiabatic potential curves for adsorbed atoms on the atomically flat Si(001) surface, however, the maximum external electric field of $\sim$ 2.6 V/\AA \hspace{2mm} applied in their simulation is rather small compared to that in experiments using field ion microscopy (3.8 V/\AA) \cite{tsong}. We now determine activation energies for silicon atoms on Si(001) surfaces under the stronger electric fields of 3.0, 3.5, and 4.0 V/\AA. Moreover, we examine differences in activation energies and threshold values of external electric fields among silicon atoms evaporating from various Si(001) surfaces, e.g., the atomically flat surface, the surface having a step, and the atomically flat surface with an adsorbed silicon atom. We find that the threshold value for field evaporation of the atom on the atomically flat Si(001) surface, which is between 3.0 and 3.5 V/\AA, is the lowest and that for the surface atom in the atomically flat surface, which is over 4.0 V/\AA, is the highest. These results will be explained in terms of binding energies between evaporating atoms and surfaces.

The rest of the paper is organized as follows: in Sec.~\ref{sec:cm}, we describe briefly the computational details of our calculations. The results are presented and discussed in Sec.~\ref{sec:rd}. We conclude our findings in Sec.~\ref{sec:concl}.

\section{Computational details}
\label{sec:cm}
\subsection{Method}
Our first-principles molecular-dynamics simulations are based on the real-space finite-difference method \cite{chelikowsky} with incorporation of the timesaving double-grid technique \cite{tsdg}. Compared with the plane-wave approach, the real-space finite-difference method is much simpler to implement while keeping a high degree of accuracy. Moreover, the real-space calculations eliminate the serious drawbacks of the conventional plane-wave approach such as its inability to describe strictly nonperiodic systems: in the case of simulations under external electric fields, the periodic boundary condition gives rise to saw-tooth potential, which sometimes leads to numerical instability during the self-consistent iteration, while the real-space finite-difference method can exactly determine the potential by the external electric field as a boundary condition and is free from involving the saw-tooth potential. We adopt the nine-point finite-difference formula for the derivative arising from the kinetic-energy operator of the Kohn-Sham equation in the density functional theory \cite{hks}. We take a cutoff energy of 23 Ry, which corresponds to a grid spacing of 0.65 a.u., and a higher cutoff energy of 210 Ry in the vicinity of nuclei with the augmentation of double-grid points \cite{tsdg}. The norm conserving pseudopotential \cite{pseudop,kobayashi} is employed in a Kleinman-Bylander nonlocal form \cite{kleinman-bylander}. Exchange-correlation effects are treated with local density approximation \cite{lda}.

\subsection{Models}
Figure \ref{fig:1} shows the top views of the Si(001) surfaces calculated here. We employ a technique that involves the use of a supercell whose size is chosen as $L_x$=43.52 a.u., $L_y$=14.51 a.u. and $L_z$=32.64 a.u., where $L_x$, $L_y$ and $L_z$ are the lengths of the supercell in the $x$, $y$ and $z$ directions, respectively. Here the direction perpendicular to the surface was chosen as the $z$ direction. In order to eliminate completely unfavorable effects of atoms in neighbor cells which are artificially repeated in the case of the periodic boundary condition, we impose the {\it nonperiodic} boundary condition of vanishing wave functions out of the supercell in the $z$ direction, while we adopt the periodic boundary condition in the $x$ and $y$ directions. The supercell contains five silicon layers, the lowest of which is terminated by hydrogen atoms (i.e., the thin film model). The atoms in the three topmost silicon layers are fully optimized by the first-principles molecular-dynamics simulation. In the cases of an atomically flat surface [terrace model in Fig. \ref{fig:1} (a)] and a surface having a step [$S_A$ step model in Fig. \ref{fig:1} (b)], we employ previously reported atomic configurations \cite{si22}. On the other hand, for the atomically flat surface where an atom is adsorbed [adatom models in Figs. \ref{fig:1} (c) and (d)], we determine the minimum-energy atomic configuration. The atom is initially placed at each of the sites of A, B, C, D, E, F, and G in Fig. \ref{fig:1} with an appropriate distance from the surface, and then the forces on the atoms are relieved during the structural optimization. The total energies are collected in Table \ref{tbl:1}. The most stable site is found to be site E in Fig. \ref{fig:1} (d), and hereafter we use this configuration as the adatom model.
\begin{table}
\caption{Total energies as the atom is displaced at sites A-G. The zero of total energy is chosen to be that at site E.}
\begin{tabular}{c|c}
\hline \hline 
Site & Total energy (eV) \\ \hline
A & 1.42  \\
B & 1.71  \\
C & 0.98  \\
D & 1.51  \\
E & 0.00  \\
F & 1.14  \\
G & 1.28  \\
\hline \hline
\end{tabular}
\label{tbl:1}
\end{table}

\begin{figure}[htb]
\vspace*{12.78cm}
\caption{Top views of the three topmost layers of Si(001). (a) the atomically flat surface (terrace model), (b) the surface having a $S_A$ step ($S_A$ step model), (c) and (d) the surfaces where an atom is adsorbed (adatom models). Closed circles, opened circles with solid curves, and opened circles with dotted curves represent atoms on the top, second, and third layers, respectively. Large (small) circles represent the upper (lower) atoms in buckled dimers.}
\label{fig:1}
\end{figure}

\section{Results and discussion}
\label{sec:rd}
\subsection{Field evaporation of surface atoms}
\label{sec:3A}
We first apply an external electric field $F$ of 3.0 V/\AA \hspace{2mm} along the $z$ direction and find that no surface atoms evaporate in any models. When the external electric field $F$ is increased to 3.5 V/\AA, the adsorbed atom in the adatom model evaporates [see Fig. \ref{fig:2} (a)]. On the basis of these results, it is evident that the threshold value of the external electric field for evaporation is 3.0-3.5 V/\AA. This value is in agreement with the experimental data of 3.8 V/\AA \hspace{2mm} obtained using a field ion microscope. In the case of $F$=4.0 V/\AA, the two upper atoms of the buckled dimers located at the step edges simultaneously evaporate in the $S_A$ step model [see Fig. \ref{fig:2} (b)] \cite{comment1}. The surface atoms in the $S_A$ step model are preferentially removed from the step edges. No surface atoms in the terrace model evaporate at all below $F$=4.0 V/\AA.
\begin{figure}[htb]
\vspace*{7.54cm}
\caption{Isosurfaces of the charge density for (a) the adatom model under $F$=3.5 V/\AA \hspace{2mm} and (b) the $S_A$ step model under $F$=4.0 V/\AA. Light and dark balls represent the surface atom and the atoms in the silicon surface, respectively.}
\label{fig:2}
\end{figure}

\subsection{Activation energy}
One of the potentially important applications of the field-evaporation process is the direct determination of the binding strengths of the surface atoms from the external electric field required for their removal. We now evaluate the activation energies of surface atoms for field evaporation during the lifting up of the surface atoms. Table \ref{tbl:2} shows the activation energies for various surface models \cite{comment2}. The activation energies in the absence of the external electric field, which are equal to the binding energies between evaporating atoms and surfaces, are observed to be 6.03, 5.86, and 5.04 eV for the terrace, $S_A$ step, and adatom model, respectively. These energies become lower as the external electric fields are increased. The threshold value of the external electric field varies with the bond strength of the surface atom, and the rate constants of field evaporation depend on the activation energies according to the Arrhenius formula \cite{Arrhenius}; surface atoms in both the $S_A$ step model and the adatom model are easily removed compared with the surface atom in the terrace model. This situation may lead to the development of technology to create atomically flat surfaces by applying the external electric fields to surfaces and adjusting the temperature.
\begin{table}
\caption{Activation energies (in eV) \cite{comment2}. The values at zero field are the binding energies between the lifted-up atoms and the surfaces.}
\begin{tabular}{c|cccc}
\hline \hline
           &0.0(V/\AA)&3.0(V/\AA)&3.5(V/\AA)&4.0(V/\AA) \\ \hline
terrace    &  6.03    &   1.45   &    0.52  &    0.24   \\
$S_A$ step &  5.86    &   1.05   &    0.13  &    0.00   \\
adatom     &  5.04    &   0.09   &    0.00  &    0.00   \\
\hline \hline
\end{tabular}
\label{tbl:2}
\end{table}

\begin{figure}[htb]
\vspace*{11.94cm}
\caption{Electronic density shift $\rho ({\bf r},F)-\rho ({\bf r},F=0)$ at $F$=3.5 V/\AA \hspace{2mm} represented on the cross section of the ribbon containing the thick line A-B. The contour spacing is 6.7 electron/supercell. Solid (dotted) curves represent nonnegative (negative) values. The large and small balls indicate the atomic positions on and above the cross section, respectively.}
\label{fig:3}
\end{figure}
\begin{figure}[htb]
\vspace*{11.96cm}
\caption{Difference in external electrostatic field between $F$=0.0 V/\AA \hspace{2mm} and $F$=3.5 V/\AA. The contour spacing is 0.5 V/\AA. The meanings of the symbols are the same as those in Fig. \ref{fig:3}.}
\label{fig:4}
\end{figure}

\subsection{Local-field enhancement}
Figure \ref{fig:3} shows the electron density shift $\rho ({\bf r},F)-\rho ({\bf r},F=0)$ due to the application of the external electric field of $F$=3.5 V/\AA \cite{comment3}. The overall charge around the surface atoms decreases, and the atom is expected to be a positive ion when it evaporates. We then calculate the external electrostatic field as the difference between the total electrostatic field in the presence of an external electric field and that in the absence of it, $d(V_{eff}({\bf r},F)-V_{eff}({\bf r},F=0))/dz$, where $V_{eff}$ is the sum of the external, Hartree, and exchange-correlation potential functions. Figure \ref{fig:4} shows the counter plots of the difference in external electrostatic field \cite{comment3}. The expulsion of the external electronic field from inside the surface can be clearly recognized in all models. However the local-field enhancement occurs not around the evaporating atom but above it, and no further significant difference in external electrostatic field around the evaporating atom is observed. Therefore, we cannot conclude that the local field around the evaporating atom plays a crucial role in the situation mentioned above in Sec. \ref{sec:3A}.

\section{Conclusion}
\label{sec:concl}
We have studied the field-evaporation process of surface atoms by using various surface models. The threshold value of the external electric field for evaporation of the surface atom on the atomically flat Si(001) surface, which is between 3.0 and 3.5 V/\AA \hspace{2mm} and is in agreement with the experimental result of 3.8 V/\AA, is the lowest among the models in this study, and no surface atoms of the atomically flat Si(001) surface evaporate below the external electric field of 4.0 V/\AA. When the external electric field is applied to the surface having a $S_A$ step, the atoms which are located at the step edges are preferentially removed. The threshold value of the external electric field obviously depends on the binding energy between evaporating atoms and surfaces. Moreover, as the external electric field increases, the activation energy for field evaporation becomes lower. On the basis of the discussions above, we conclude that the field-evaporation process can be applied to surface-flattening techniques \cite{iwasaki} or atomic-scale manipulation methods \cite{eigler1,lyo,eigler2} by adjusting the strength of the external electric field and the temperature.

\section*{Acknowledgements}
This research was partially supported by the Ministry of Education, Culture, Sports, Science and Technology, Grant-in-Aid for COE Research (Grant No. 08CE2004) and Young Scientists (B) (Grant No. 14750022). The numerical calculation was carried out by the computer facilities at the Institute for Solid State Physics at the University of Tokyo, and Okazaki National Institute.


\begin{references}
\bibitem{comment1}
There are two $S_A$ step edges in the supercell, because we imposed periodic boundary condition in the directions parallel to the surface. Each of two upper atoms of the dimers is located slightly different situation, however deferences in activation energies for the field evaporation between them are negligibly small.
\bibitem{comment2}
For the $S_A$ step model, we depict the activation energies of the center atom in Fig. \ref{fig:1} (b).
\end{references}
\end{document}